# Quasiparticle scattering off phase boundaries in epitaxial graphene

A Mahmood*, P Mallet and J-Y Veuillen

Institut Néel, CNRS-UJF, Boîte Postale 166, 38042 Grenoble, France

* Corresponding author

E-mails: ather.mahmood@grenoble.cnrs.fr, pierre.mallet@grenoble.cnrs.fr, jean-yves.veuillen@grenoble.cnrs.fr

**Abstract**

We investigate the electronic structure of terraces of single layer graphene (SLG) by scanning tunneling microscopy (STM) on samples grown by thermal decomposition of 6H-SiC(0001) crystals in ultra-high vacuum. We focus on the perturbations of the local density of states (LDOS) in the vicinity of edges of SLG terraces. Armchair edges are found to favour intervalley quasiparticle scattering, leading to the ($\sqrt{3} \times \sqrt{3}$)$R30°$ LDOS superstructure already reported for graphite edges and more recently for SLG on SiC(0001). Using Fourier transform of LDOS images, we demonstrate that the intrinsic doping of SLG is responsible for a LDOS pattern at the Fermi energy which is more complex than for neutral graphene or graphite, since it combines local ($\sqrt{3} \times \sqrt{3}$)$R30°$ superstructure and long range beating modulation. Although these features were already reported by Yang et al. Nanoletters 10, 943 (2010), we propose here an alternative interpretation based on simple arguments classically used to describe standing wave patterns in standard two-dimensional systems. Finally, we discuss the absence of intervalley scattering off other typical boundaries: zig-zag edges and SLG/bilayer graphene junctions.

## 1. Introduction

Although a subject of theoretical studies for more than half a century [1], the discovery of single sheets of graphene that remain stable in ambient conditions [2] has led to the observation of fascinating properties of graphene [3]. Considered as a condensed matter analogue for quantum electrodynamics in (2+1) dimension, graphene is also of particular interest for nanoelectronics due to high mobilities and near-ballistic transport at room temperature [4]. The manifestation of phenomena like anomalous Quantum Hall Effect [3][5] and weak (anti-) localization [6][7] render it important from a perspective of study of fundamental physics. Most of these phenomena arise from the peculiar energy band structure of graphene at low energy.

The quasiparticle states near the Fermi level are described by a valley-index that specifies the valley at $K$ and $K'$ points of the Brillouin zone (BZ) to which the state belongs. Further, the specific symmetry present in electronic states of each valley is represented by a pseudospin that is a measure of the relative wavefunction amplitude on each sublattice A and B of the graphene unit cell. These quasiparticle symmetry properties present in perfect graphene layers may be violated due to the presence of short range potentials that arise from the atomic scale defects present in the graphene lattice and may cause them to scatter and interfere.

The presence of graphene edges is also responsible for symmetry breaking and depending on the edge orientation incident quasiparticle states may scatter differently. It has been proposed that scattering by zigzag edge produces intravalley scattering whereas armchair edge results into intervalley scattering [8]. These scattering events are responsible for quasiparticle interference (QI) effects which in graphene may be impacted by pseudospin conservation [8]. Scattering effects in graphene have a profound impact on electron transport properties with suppression or enhancement of weak localization according to the presence or absence of time-reversal symmetry, respectively [7]. It is important to characterize directly the scattering mechanisms of the different kinds of defects in a given sample to get a better understanding of their transport properties. We present here an analysis of the scattering properties of the most common defects of graphene layers epitaxially grown on SiC; namely phase boundaries [9][10][11][12]. These include zigzag and armchair edges and monolayer/bilayer junctions.

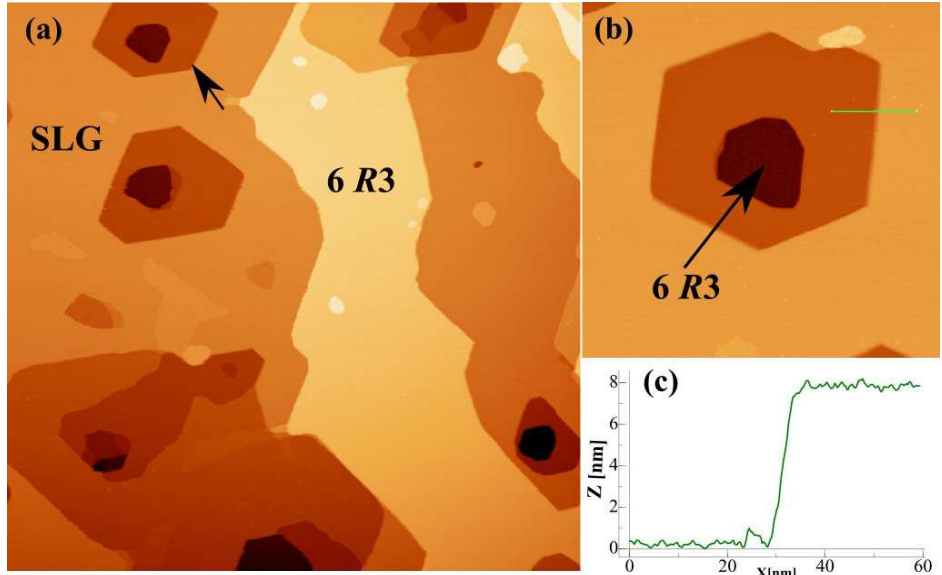

Figure 1: (a) Large scale (800 x 800 nm²) UHV-STM image of a partially graphitized SiC(0001) surface covered by single-layer graphene (SLG), with some patches of $6R3$ reconstructions found either on the surface or at the bottom of the pits (sample bias, $V_t$ = -0.5V, tunnelling current, $I_t$=0.2nA). Note the surface features such as flat terraces, pits and edges. (b) A zoom of the pit marked by an arrow in Panel (a) of a $6R3$ region surrounded by SLG terraces. (c) A profile of the step between two SLG terraces in (b) indicates a step height equal to 3 bilayers of SiC.

Although, graphene edge has been intensively studied theoretically [8][13][14][15][16] most of the experimental work is limited to either transport studies of nanoribbons [17] or local probe studies of steps on HOPG [16][18]. Recently, scanning tunnelling microscopy (STM) images of edges of graphene layers grown epitaxially on Si face of SiC crystal have been reported [19][20][21].

The present study is dedicated to the same system. We use STM images of the interference patterns of electron waves near phase boundaries in graphene to reveal the scattering properties of these defects. Scattering at armchair edges is analyzed in much the same way as has previously been reported near the edges in metal surfaces [22][23]. Our analysis is somewhat different from studies by Yang *et al*.[21], and goes beyond the one of reference [19]. Considering the interference pattern of electronic waves near the edges as a fine fingerprint of scattering events the role of edge geometry is better understood with strong implications for electron transport in the vicinity of edges.

## 2. Experimental details

Graphitized samples were fabricated in a home built ultra-high vacuum (UHV) system maintained at a base pressure of ~$10^{-10}$ mbar during growth, following the standard procedure[9]·[24]. Wafers of 6H-SiC (0001) polytype were cleaned by ultrasonication in ethanol and acetone for dust and grease removal followed by out gassing at 600°C for several hours in UHV. The substrate was then annealed at 850°C under a Si flux for oxide removal and preparation of a Si rich surface prior to graphitization. Upon increasing the temperature from 900°C to 1100°C during further annealing steps, ($\sqrt{3}\times\sqrt{3}$) $R30°$ and ($6\sqrt{3}\times6\sqrt{3}$) $R30°$ phases covering the SiC surface were obtained. Finally, the surface was graphitized by heating the substrate between 1300°C and 1350°C, and by adjusting the parameters such as temperature and annealing time, the number of graphene layers could be controlled. The phase changes and graphitic growth was fully characterized by in-situ low energy electron diffraction (LEED) and Auger electron spectroscopy (details not shown). As elaborated in the next section, the growth parameters were chosen to obtain single layer graphene (SLG) coverage, while the presence of patches of the carbon rich ($6\sqrt{3}\times6\sqrt{3}$) $R30°$ phase (referred as buffer layer or 6R3 in the remaining text) had to be ensured in order to study the electronic interactions at the junction of SLG and buffer layer. This coverage corresponded to a C/Si peak ratio ~1 in Auger electron spectroscopy. The formation, nature and electronic structure and the interaction of the buffer layer with the underlying SiC substrate has been extensively studied [25][26]. It may be noted that the buffer layer has a honeycomb arrangement of C atoms with C-C bond length that is similar to graphene but lacks the π-bands that renders it insulating in nature [27]. Further, the buffer layer exhibits an apparent SiC-6×6 superstructure on STM images[9][10][11][28]. The sample surface was imaged with a home built UHV STM system where the sample could be transferred without exposing it to air. Topographic STM images were obtained at room temperature in constant current mode with mechanically cut PtIr tips and with a bias applied to the sample.

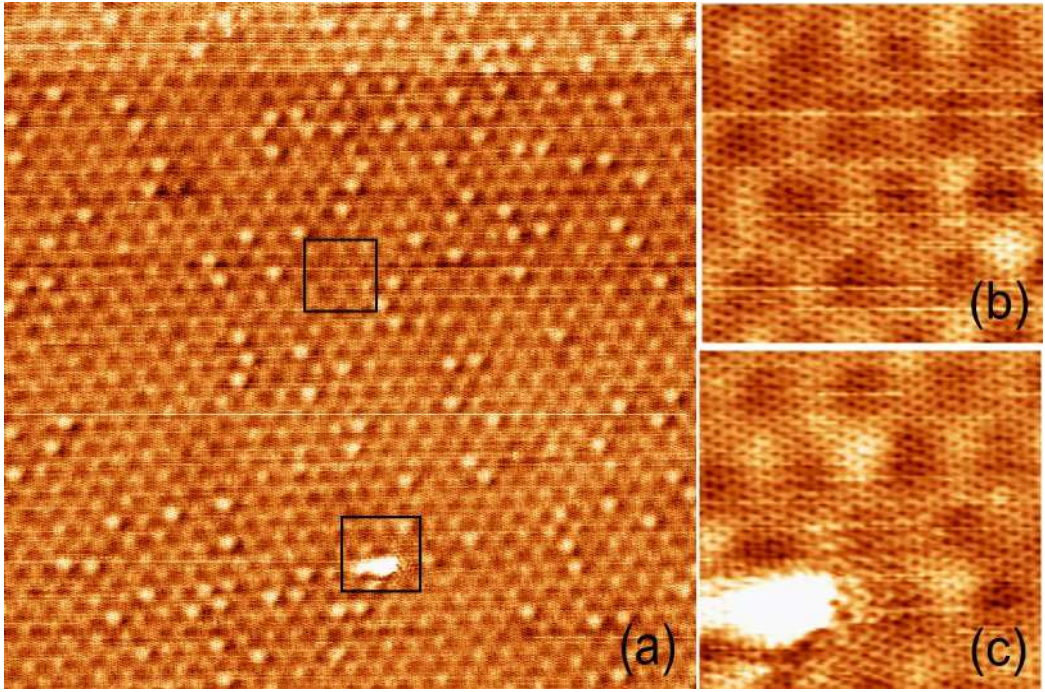

Figure 2: (a) A 50×50 nm² low bias ($V_t$=-30mV, $I_t$=0.2nA) STM image of SLG region with underlying 6R3 interface visible. Direct zoom over the regions marked by black boxes in (a) show atomically resolved image of SLG (panel b; size: 6.5 x 6.5 nm²), and electronic modulations visible around the lattice defect (panel c; size: 6.5 x 6.5 nm²)

### 3. Topography of graphene samples

Figure 1(a) presents a large area STM topographic image of the sample surface showing various features such as steps, terraces and pits. These topographic features conform well to the earlier reported STM studies on such samples[9][10][11][12]. It may be noted that a significant area of the sample surface is covered by SLG with patches of 6R3 found at the bottom of the pits and sometimes at the surface. Figure 1(b) shows a zoomed in STM image of the upper left pit shown in panel (a). This image reveals the two representative phases present on different terraces. The 6R3 reconstructed surface lies at the bottom of the pit whereas the two terraces comprise of SLG. A step height of 7.5 Å exists at the junction between two SLG terraces as shown by the section profile (Panel c). The step height equals the distance between 3 bilayers of SiC and is a useful tool for determining the graphene layer thickness. For brevity, we skip the details of graphene layer determination method employed during this study[9][10][11].

STM images with higher resolution are able to elucidate the surface structure at atomic scale and to simultaneously depict the defects present over it and the resulting disturbances in the electronic structure. Figure 2(a) is an STM image with atomic resolution of a large SLG terrace. This image is dominated by the apparent SiC-6×6 superstructure which confirms the underlying buffer layer [9][10]. A direct zoom over this image (top square) reveals the honeycomb pattern of graphene lattice (Figure 2(b)). The bottom square in Figure 2(a) shows the presence of an extrinsic impurity on the sample surface. Panel (c) shows an additional superstructure surrounding the defect, which is commensurate with the graphene lattice. This superstructure which extends a few nm from the defect is known to have an electronic rather than a topographic origin.

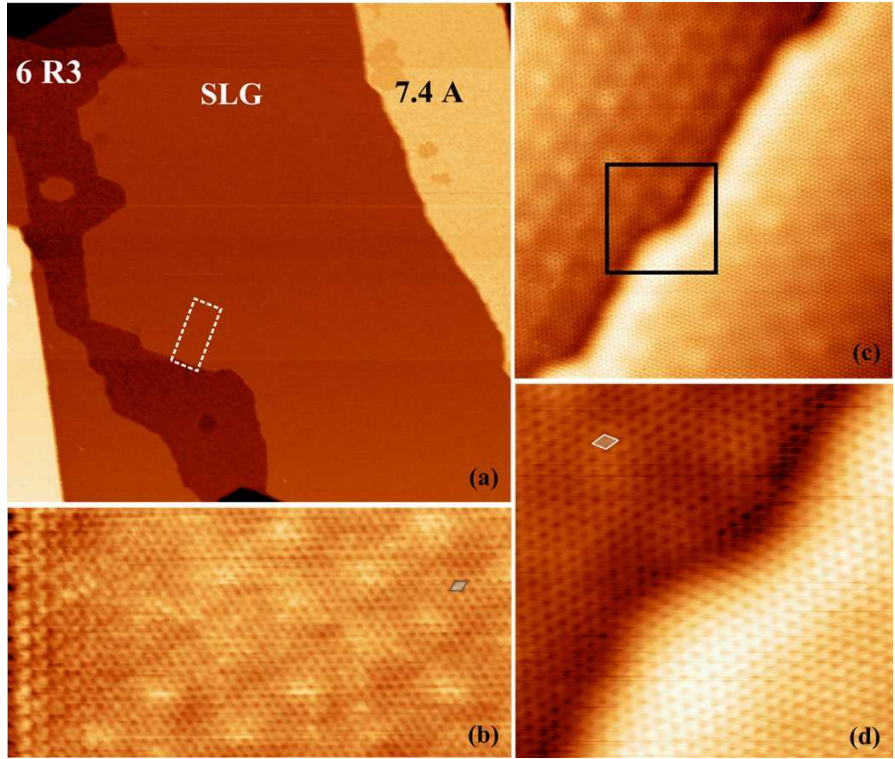

Figure 3: (a) 300 x 300 nm² STM image of a SLG terrace lying between two 6R3 regions. Clearly visible step edges with zigzag, armchair, and mixed configurations terminate at the junction of SLG and 6 $R3$ zone ($V_t$=-1.0 V). (b) 11 x 5.6 nm² low bias STM image ($V_t$=-0.15V) showing quantum interferences at a regular armchair edge of the SLG terrace (white box in Panel a). (c) 18 x 18 nm² STM topography showing adjoining SLG and BLG terraces. A zoom in Panel d (7 x 7 nm²) indicates the continuity of atomic arrangement in the surface graphene layers at the junction and the absence of $R3$ superstructures ($V_t$=-0.1 V for (c) and (d)). The graphene unit cell is indicated in Panels (b) and (d) by small diamonds.

Indeed, as reported earlier for SLG and BLG on SiC(0001) [9][29][30][31] and HOPG[15][32][33], these are identified as ($\sqrt{3}\times\sqrt{3}$) $R$ 30° superstructures ($R3$ modulation in the following) arising from impurity induced QI effects. It has been argued that the local perturbation caused by the defect results into scattering by impurities between $k$ states present in inequivalent Dirac cones at $K$ and $K'$ points, also referred as intervalley scattering [31][32][33]. Following a similar argument in the next section we will show that step edges at junction of buffer layer and SLG act as scatters of quasiparticles and lead to periodic oscillations reminiscent of Friedel oscillations in metals [22][34][35][36].

## 4. Quantum interference effects

We now focus on the step edges at the junction of buffer layer and SLG, and between SLG and BLG. A direct junction between buffer layer and BLG is very rare and was not found at sufficiently large terrace size to provide optimum resolution for QI imaging. Figure 3 (a) shows an STM topography scan of a 6$R3$ zone next to a SLG terrace (step height of 1.7 Å). Panel (b) shows a zoom-in of the selected junction, acquired after rotation and moving the tip to bring the insulating buffer zone just on one side of the scanned area. As can be readily identified from the image, the step edge comprises a regular armchair arrangement of atoms. Following Tersoff and Hamann [37], an atomically resolved image at low bias can be approximated as a map of the LDOS close to Fermi energy. QIs associated to quasiparticle scattering off the step edge at surface step result into LDOS pattern that is parallel to the step edge and extends until a distance of around 10 lattice constants from the edge. Beyond that, the hexagonal graphene lattice remains intact. A markedly different result is found for the armchair SLG/BLG junction (Figure 3(c) and 3(d). As already quoted [10][38], the top graphene layer is continuous over the junction, which is evident in Figure 3 (d). More importantly, the $R3$ superstructure develops neither on SLG nor on BLG at the boundary. Since intervalley scattering remains kinematically allowed at the SLG/BLG armchair junction, the lack of $R3$ superstructure suggests that the scattering potential of this junction do not possess Fourier components with large wavevectors [39], i.e. the potential is slowly varying ("smooth") in space.

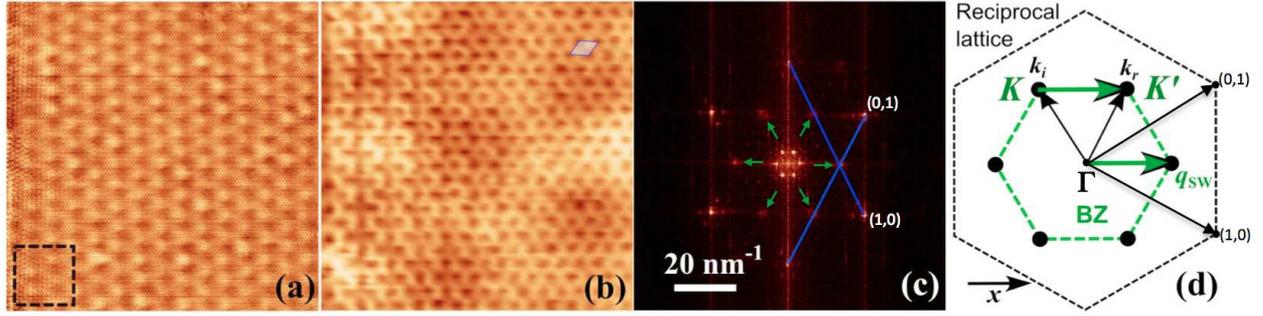

Figure 4: (a) 24 x 24 nm² STM image of an armchair edge at low bias ($V_t$=-30 mV) with direct zoom-in in Panel b (4 x4 nm²) showing the formation of standing wave patterns near the edge. The graphene unit cell is shown in the upper-right part of Panel b. (c) Fourier Transform (FT) of the image in panel (a). The green arrows point towards intensity peaks related to intervalley scattering. (d) The schematic presents the first Brillouin Zone (BZ) and the Fermi surface reduced to $K,K'$ points of the BZ: $q_{sw}$ represents the resultant standing wavevector from the interference of states $k_i$ and $k_r$ ($x$ is a unitary vector perpendicular to the step edge). Two of six first-order spots of the reciprocal lattice are labelled (1,0) and (0,1) on Panels (c) and (d).

In the following, we focus on the QIs generated close to armchair edges such as Figure 3(b). The associated LDOS pattern has already been reported by several groups [19][20][21]. Our main purpose is to present a very simple interpretation of such pattern, based on intervalley scattering, somehow different from the arguments given in reference [21]. The STM images in Figure 4(a) and (b) show with atomic resolution a junction between SLG and 6R3 area (image taken up to the step edge) at a sample bias of -30 mV. Panel b is a clear demonstration of the LDOS modulations pattern at the Fermi level due to QIs at the step edge. The later can be represented by a sharp potential step that can scatter an incident electron state ($k_i$) into a reflected wave ($k_r$) which interferes with the incident one, thus generating a standing wave (SW) of electron density. The sum of these SWs for all possible incident vectors at the Fermi surface (FS) will result in modulation of the LDOS at the Fermi energy [35]. As mentioned earlier, low bias STM images can be interpreted as LDOS maps at the Fermi level [37]. A 2-D Fourier Transform (FT) of the STM image can thus extract the information about the wave vectors of electrons at the Fermi energy that are confined on the FS contour [34].

Panel (c) shows the corresponding FT of the STM image in (a) where the six outer spots correspond to graphene lattice. The six bright inner spots are related to the SiC-6x6 superstructure showing up in (a) and related to the buffer layer, as explained in section 3. These spots are not relevant in the following. We now focus on the 6 spots of panel (c) indicated by the green arrows. These spots are rotated by 30° to the graphene lattice and their positions correspond to the six corners ($K$ and $K'$ points) of the first BZ in reciprocal space. Thus, the interference patterns in this STM image display a local $R3$ superstructure with respect to the graphene lattice. The occurrence of $R3$ structure can be understood by considering the FS of graphene. If we neglect the electron doping which shall be considered later, the FS is reduced to $K$ and $K'$ points of the BZ, or equivalently to six points due to symmetry of the reciprocal lattice (Figure 4 d). The Fermi wave vectors, of modulus $k_F = \Gamma K = 4\pi/3a$ with $a$ the lattice parameter, are directed towards the corners of BZ (schematic in Panel d). The component of the wavevector of the incident state parallel to the step edge ($k_{//}$) should be conserved upon scattering by an extended linear (translational invariant) step edge [8]. Thus intervalley scattering events are kinematically allowed at armchair edges. A large momentum scattering between two states $k_i$ and $k_r$ lying on neighboring $K$ and $K'$ points corresponds to a momentum change of $q_{SW} = k_F x$. This interaction results into a modulation of the LDOS with a wavelength of $\lambda_F=3a/2$ with a SW pattern parallel to the step.

Interestingly, the FT-LDOS shown in Figure 4 (c) exhibits spots at $\pm k_F$ along the $x$-axis, but also at other $K,K'$ points out of the $x$-axis. It is surprising since the intervalley scattering processes in Panel d only involves a modulation of wavevector parallel to $x$. As detailed in the supplementary information, such extra spots at $K,K'$ points are the replica of the spots at $\pm k_F x$, obtained by translations of vectors of the graphene reciprocal lattice. Such replica are expected due to the Bloch nature of quasiparticles, as already reported in conventional 2D systems [40][41]. Consequently, the LDOS perturbation close to armchair edges is not purely one-dimensional, and rather forms a local $R3$ pattern, despite the 1D symmetry of the edge.

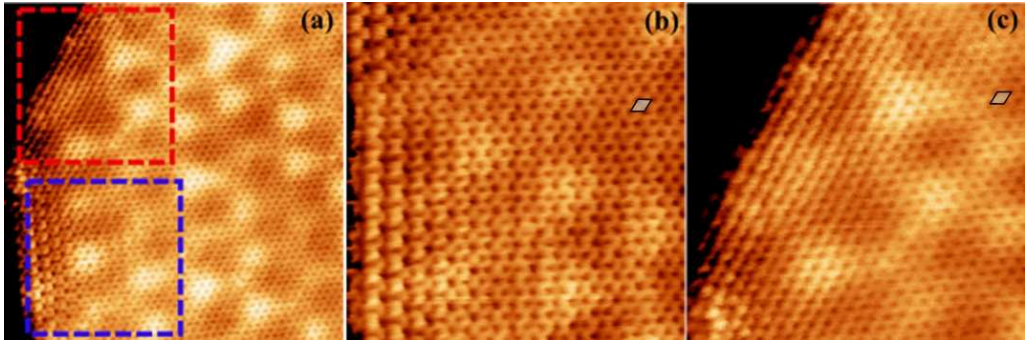

Figure 5: Comparison of electronic modulations at armchair (blue box) and zigzag edges (red box) is shown in a 10 x 10 nm² STM image ($V_t$ = + 100 mV) in (a). The zoomed-in (5.5 x 5.5 nm²) images of perturbations near armchair (b) and zig-zag edge (c) clearly demonstrates the differences in superstructure formation. Away from the edges, small diamonds in Panels (b) and (c) mark the graphene unit cell.

It should be emphasized that the observed standing waves with an $R3$ periodicity has only been noticed in the case of scattering off the armchair edges. Figure 5 (a) clearly shows a terrace, where the two sides of the steps are recognized as armchair (Panel b) and zig-zag (Panel c). As expected, zig-zag edge does not show any $R3$ pattern, although modulated LDOS pattern is clearly seen with periodicity corresponding to graphene lattice. The absence of $R3$ modulation is explained if we consider that the BZ for zigzag edge is rotated by 30° and that the vector component $k_{//}$ of incident quasiparticle parallel to the edge is conserved. In this situation (zigzag edge), intervalley scattering ($K/K'$) is kinematically forbidden. The scattering event will however generate wavepatterns with a wave number equal to a reciprocal lattice vector. It should be noted that these lines are not a consequence of resolution change since the honeycomb pattern is apparently present away from the edge.

The case of quasiparticle interference presented above is only a simple picture where the Dirac point is assumed to be at Fermi energy level in which case the FS contour is reduced to two points, $K$ and $K'$. Actually, the FS is made of a circular pocket of radius $q_F$ at points $K$ and $K'$ resulting from the shift of the Dirac point by 0.45eV below the Fermi level for graphene on SiC (0001) [25][42]. Figure 6(a) shows a large sized STM image (24 nm) of a terrace with an armchair edge, taken at a low bias of -30mV and panel b is the FT of this image. A 24 nm scan size with enough points is optimal to enhance the resolution in reciprocal space, which facilitates easily discernible spots in FT with sufficient atomic resolution of the topography image. A zoom on $K,K'$ points in the FT indicates that each spot is split in two spots separated by ~0.22Å⁻¹ ± 0.03 Å⁻¹ (see inset of Figure 6 (b))

As quoted previously, and shown in Figure 6(c), intervalley scattering mixes an incident state $k_i$ around $K'$ with reflected (scattered) states $k_r$ around $K$. Owing to the conservation of $k_{//}$ (illustrated by the continuous line in Figure 6-c), there may be 2 scattered (reflected) states around K. However, only the scattered states labelled $k_r$ corresponds to a reflected state which propagates away from the step (the groups velocity of the incident and scattered states should have opposite component perpendicular to the step). Hence, from this incident state $k_i$ and reflected state $k_r$ a standing wave with wavevector $q=\pm(k_r-k_i)$ is generated. The positive component along the $x$ axis in Figure 6-c is $q=(\Gamma K+2q_F\cos\theta)x$ ($\theta$ is the incidence angle of the incoming wave; $x$ is a unitary vector perpendicular to the step). Considering an incident wave in the $K$ valley, a similar reasoning lead to a SW pattern with wavevector $q'=(\Gamma K-2q_F\cos\theta)x$. The total signal due to intervalley scattering results from summing up contributions from SW pattern with wavevectors $q$ and $q'$ over all values of $\theta$. The backscattering events ($\theta$~0°) will dominate this sum since these processes correspond to the largest joint density of states (JDOS)[22][29][34][36]. Thus the main Fourier components of the scattering pattern at an armchair edge will be located at $(\Gamma K\pm2q_F)x$, with also replica of these two peaks at other $K,K'$ points as discussed above (see supplementary information). This explains the 2 spots in the inset of Figure 6(b). In agreement with this interpretation, the measured value of the splitting of the doublet (0.22Å⁻¹ ±0.03Å⁻¹) is close to the estimated $4q_F$ value (0.24Å⁻¹) with $q_F$=0.06Å⁻¹ [31].

In real space images, the superposition of the two SWs with wavevector $\Gamma K\pm2q_F$ should give rise to a beating modulation with an apparent periodicity $\pi/2q_F$ ~ 29 Å, as shown in supplementary information. Following Ref. [21], this beating structure is revealed in the inverse FT (IFT) map (Figure 6 (d)-and the corresponding profile of Figure 6 (e)) - obtained by filtering out the FT spots other than the doublets at the $K/K'$ points in Panel b. This simple consideration based on the shape of the real FS (Figure 6(c)) explains two important aspects: Firstly, the short range modulation (SW at wavevector ≈ΓK), known as the $R3$ pattern implying a modulation with wavelength $\lambda=3a/2$; secondly, the apparent long range modulation, which is of beating origin, of wavevector $4q_F$ and a wavelength of $\pi/2q_F$. This long range modulation originating from intervalley scattering processes is thus different from a possible modulation of wavevector $2q_F$ which could arise due to intravalley backscattering [31]. Note that theory predicts that the latter processes are absent close to armchair edges due to kinematical and pseudospin considerations [8].

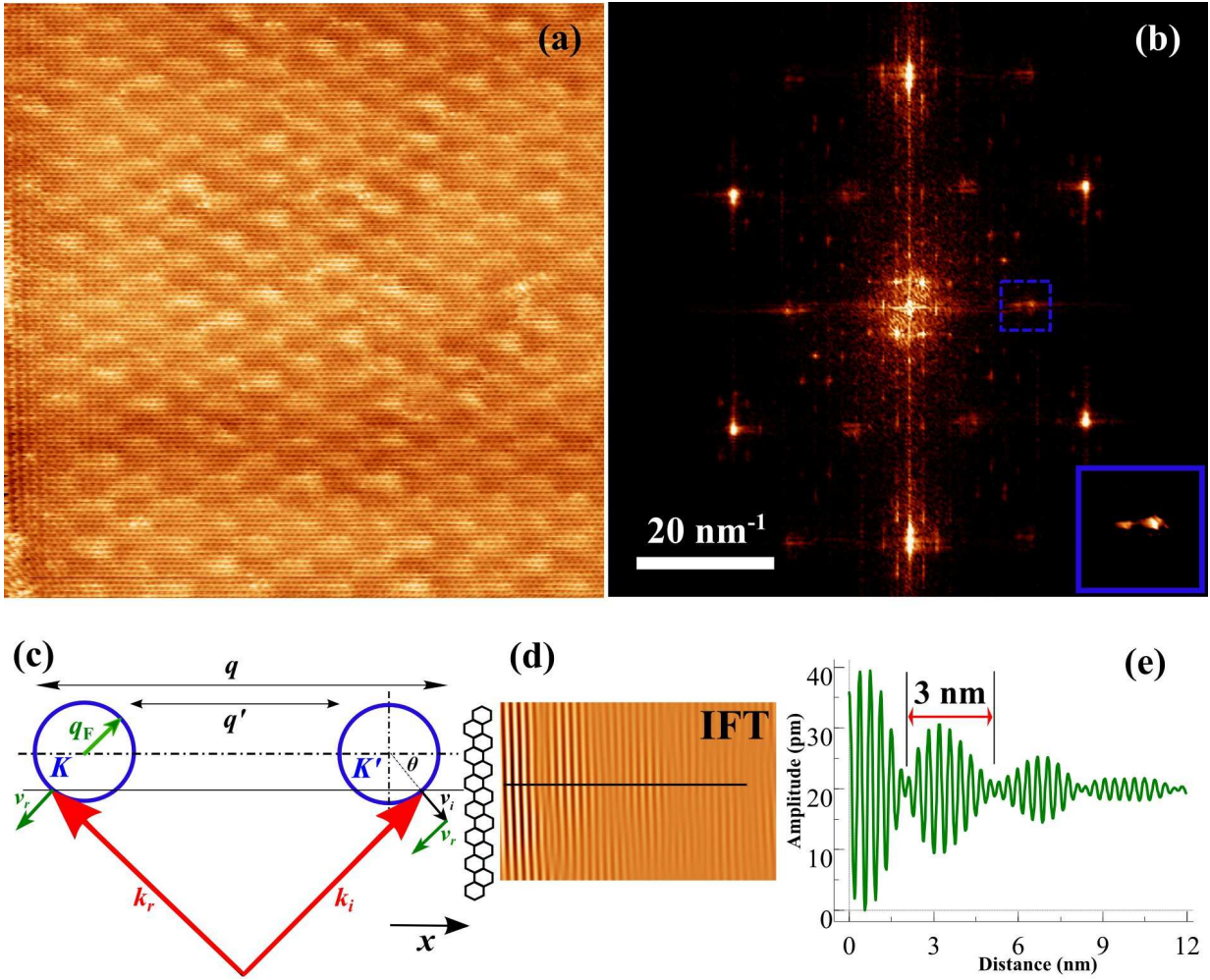

Figure 6: Quantum interference off an armchair edge for doped graphene. (a) 24 x 24 nm² STM topography image of an armchair terminated step edge, acquired at a low bias ($V_t = -30$ mV) and the corresponding FT image in (b). The splitting of the intensity at one $K$ point is clearly seen as a doublet in the inset. (c) Schematic explaining intervalley scattering processes in doped graphene. An incident state $\mathbf{k_i}$ (with incidence angle $\theta$ and velocity $\mathbf{v_i}$) is scattered into the reflected state $\mathbf{k_r}$ (with velocity $\mathbf{v_r}$) at the step edge. The component of the wavevector parallel to the step edge is conserved, but the component of velocity perpendicular to the step edge is reversed. $\mathbf{q}$ and $\mathbf{q'}$ are the extremal wavevectors of the SWs generated at the edges. $q_F$ is the radius of the pockets of the FS at the $K$ and $K'$ points. (d) Inverse FT of the doublet near $K/K'$ (inset of (b)) which shows the decaying beating pattern of period 3 nm as displayed by the line profile in (e).

## 5. Conclusion

We have investigated the quasiparticle scattering by graphene edges in partially graphitized surface of SiC (0001) samples by means of scanning tunneling microscopy in UHV. We observe the formation of R3 superstructure in the vicinity of regular armchair edge geometry and attribute its presence to the QI effects arising from intervalley scattering between $K$ and $K'$ valleys. A Fourier transform analysis of low bias STM images close to armchair edges reveals a splitting of the intensity peaks associated to the intervalley scattering. This splitting allows a direct estimation of the doping level of epitaxial graphene. Intervalley scattering was not observed at zigzag edges for kinematical reasons. Further, the absence of intervalley scattering at armchair SLG/BL junctions is ascribed to a smooth scattering potential in the continuous surface layer. The present study of elastic scattering of quasiparticles from graphene boundaries - which are common defects in these samples - is complementary to the mesoscopic and microscopic charge transport properties in graphene since the later is affected by the conservation of chirality and pseudospin in scattering processes [43]. A study based on weak(anti-)localization effects by magneto-transport measurements has already been submitted elsewhere[44].

## Acknowledgments

This work was supported by the French ANR ("GraphSiC" Project No. ANR-07-BLAN-0161), by the Région Rhône-Alpes ("Cible07", "Cible08" and "Cluster micro-nano" programs) and by the "Fondation Nanosciences" (project RTRA "DispoGraph"). We thank F. Hiebel for valuable collaboration.